\documentclass[aps,prl,twocolumn,showpacs,floatfix,amsfonts,amssymb]{revtex4}
\usepackage{graphics}
\usepackage{psfrag}
\usepackage{placeins}
\usepackage{graphicx}
\usepackage{amssymb}
\usepackage{amsmath}
\usepackage{times}

\begin{document}

\title{Stochastic gain in population dynamics}
\author{Arne Traulsen\footnote{traulsen@theo-physik.uni-kiel.de}}
\author{Torsten R{\"o}hl}
\author{Heinz Georg Schuster}
\affiliation{Institut f{\"u}r Theoretische Physik und
Astrophysik, Christian-Albrechts Universit{\"a}t,
Olshausenstra{\ss}e 40, 24098 Kiel, Germany}
\date{ May 7th, 2004}

\begin{abstract}
We introduce an extension of the usual replicator
dynamics to adaptive learning rates.
We show that a population with a dynamic
learning rate can gain an increased
average payoff in transient phases
and can also exploit external noise,
leading the system away from the
Nash equilibrium, in a reasonance-like fashion.
The payoff versus noise curve resembles
the signal to noise ratio curve
in stochastic resonance.
Seen in this broad context, we introduce
another mechanism that
exploits fluctuations in order to improve
properties of the system. Such a mechanism
could be of particular interest in economic systems.
\end{abstract}

\maketitle
Game theory \cite{Neumann1953} describes
situations in which the success or payoff of an
individual depends on its own action as well
as on the actions of others. This paradigm can
be applied to biological systems, as evolution
through natural selection can be viewed as an
optimization process in which the fitness-landscape
changes with the state of the adaptive populations
\cite{Nowak2004}.
Evolutionary game theory focuses mainly on systems
with a single fitness function for all individuals,
which is identified with the payoff function of a game
\cite{Weibull1996,Hofbauer1998,Smith1981}.
In nature often different populations with different
ambitions interact with each other, as shoppers and
sellers \cite{Gintis2000}, attackers and defenders
\cite{Gintis2000}, or males and females
\cite{Smith1981}. Here, the payoff functions are
different for the interacting populations.
A mean-field description of such asymmetric conflicts
is given by the coupled replicator equations
\cite{Hofbauer1998,Smith1981,Schuster1981}.
These equations have a very rich dynamical behavior
and can even display Hamiltonian chaos
\cite{Sato2002a,Sato2002b}. In previous work
\cite{Smith1981, Hofbauer1998, Weibull1996}
it has been tacitly assumed that both populations
have the same adaptation mechanisms. But it seems
to be natural that different mechanisms are applied by
the interacting populations, e.g.\ different adaptation
rates. Here, we analyze such systems for the case that
both populations have slightly different adaptation
mechanisms. We assume that one population can control
its own adaptation rate. This alters the velocity when the
system is approaching the stable Nash equilibria
\cite{Holt2004} in strategy space, leading to an increased
average payoff.

In real systems fluctuations disturbing the system
are to be expected. Such disturbances can arise from
a variety of effects, e.g.\ errors of the players
\cite{Nowak1994}, deviations from a perfectly
mixed population, or immigration of individuals
with different strategy distributions. So far,
stochastic extensions to the replicator dynamics
have mainly been analyzed in the context of equilibrium
selection \cite{Foster1990,Cabrales2000}. Here, we show
that a population with adaptive learning rate can obtain an
increased payoff if these fluctuations are present. For small
noise intensities the average payoff increases,
while very large fluctuations cannot longer be exploited,
leading to a decrease of the average payoff.
This recalls the stochastic resonance effect
\cite{Benzi1981,Gammaitoni1998, Moss1989,Moss1993},
where the signal to noise ratio of a system is improved for
intermediate noise intensities. In contrast to
the usual stochastic resonance, a periodic force is
not involved here, making the mechanism more similar to
coherence resonance \cite{Pikovsky1997}.
Seen in this broader context, we introduce another mechanism that
exploits fluctuations in order to improve
the performance of the system.

We consider two adaptive species $X$ and $Y$---each
with different strategies---that are involved in a
repeated game. Both populations have different
objectives described by payoff matrices $P_x$ and $P_y$.
The fraction of individuals $x_i$ that adopt a certain strategy $i$
grows proportional to the relative payoff of the strategy $i$,
the same holds for $Y$. In the presence of noise this coevolution
can be described by the coupled replicator equations,
\begin{eqnarray}
\label{RD}
{\dot x}_i & = & x_i \eta_x \left[ \Pi^x_i -\langle \Pi^x \rangle  \right] +\xi^x_i\\
{\dot y}_i & = & y_i \eta_y \left[  \Pi^y_i -\langle \Pi^y \rangle \right] +\xi^y_i\nonumber,
\end{eqnarray}
where $\eta_x$ and $\eta_y$ are the learning
rates of the populations. We assume for simplicity
that the noise $\xi_i$ is Gaussian with autocorrelation
$\langle \xi^k_i(t)  \xi^l_j(s) \rangle = \sigma^2 \delta_{ij} \delta_{kl} \delta(t-s)$
as in \cite{Foster1990}. We also follow
\cite{Foster1990} choosing reflecting boundaries.
The payoffs are defined as
$\Pi^x_i = \left(P_x \cdot {\bf y} \right)_i$,
$\langle \Pi^x \rangle = {\bf x}^T \cdot P_x \cdot {\bf y}$,
and similarly for $y$.

We extend the usual replicator dynamics
by introducing adaptive learning rates as
\begin{equation}
\label{mrule}
\eta_x  =  1 - \tanh \left(\alpha_x \Delta \Pi  \right),
\end{equation}
where $\Delta \Pi = \langle \Pi^x \rangle - \langle \Pi^y \rangle$
is the time dependent difference between the
average payoffs of the populations
and $\alpha_x \geq 0$ is a ``perception ability''
of the population. In order to maintain the basic
features of the replicator dynamics, the learning
rate must be a positive function with
$\langle \eta \rangle=1 $, which is ensured
by Eq. (\ref{mrule}). For $\alpha_x >0$ the
population $X$ learns slower if it is currently
in a good position, otherwise it learns faster.
The value of $\alpha_x$ determines how well
a population can assess its current state.
The adaptive learning rate leads to a faster escape
from unfavourable states, while on the other
hand the population tends to remain in preferable states.
Other choices for $\eta_x$ which ensure these
properties mentioned above will not
alter our results. In the following we will
focus on a setting where only one population
has an adaptive learning rate $\eta_x$ as in
Eq.\ (\ref{mrule}).

The noise introduced above drives the system
away from the Nash equilibrium and leads for
small amplitude to a positive gain of the population
with adaptive learning rate whereas for large noise amplitudes
the fluctuations smear out the trajectories in
phase space so strongly
that they can no longer be exploited.
Hence, we expect an optimal noise effect
for intermediate values of $\sigma$.
In order to be able to compare the payoffs of
both populations we assume that the
dynamics starts from the Nash equilibrium.

As a first example, we consider
the zero sum game ``matching pennies''
\cite{Weibull1996,Schuster2002}.
Here, both players can choose
between two options $\pm 1$.
Player one wins if both players select the
same option and player two wins otherwise.
The game is described by the payoff matrices
\begin{equation}
\label{PMpayoff}
P_x=\left( \begin{array}{cc} +1 & -1 \\ -1 & +1 \\ \end{array} \right)=-P_y.
\end{equation}
The replicator equations follow from Eqs.\
(\ref{RD}) and (\ref{PMpayoff}) as
\begin{eqnarray}
\label{PMRD}
{\dot x} &=&  -2\eta_x x (2 y -1)(x-1) +\xi_x \nonumber \\
{\dot y} &=&  +2\eta_y y (2 x -1)(y-1) +\xi_y,
\end{eqnarray}
where $x=x_0$ and $y=y_0$. Let us first consider
the zero noise limit in the case $\eta_x=\eta_y=1$.
As for all zero-sum games, i.e.\ $P_x=-P_y^T$, the
system (\ref{RD}) without noise becomes
Hamiltonian and has a constant of motion
\cite{Hofbauer1996}. Here, the constant is given by
$H(x,y) = -2 \ln\left[x(1-x)\right]-2 \ln\left[y(1-y)\right]$.
The trajectories oscillate around the Nash equilibrium at
$x=y=1/2$. $H(x,y)$ is connected to the temporal
integral of the average payoff
$\langle \Pi_x\rangle =\left({\bf x}^t\right)^T \cdot P_x \cdot {\bf y}^t $
during a period with $\langle \Pi^x \rangle >0$,
\begin{eqnarray}
\label{analyt_int}
 \int_{t_0}^{t_1} \langle \Pi^x \rangle dt
 =  -\frac{H\left(x_0,\frac{1}{2} \right) - H\left(\frac{1}{2},\frac{1}{2} \right)}{4},
\end{eqnarray}
where $(x,y)=(x_0,\frac{1}{2})$ at $t_0$
and $(x,y)=(\frac{1}{2},x_0)$ at $t_1$.

\begin{figure}[htbp]
\psfrag{x}{$x$}
\psfrag{y}[][][1][180]{$y$}
\psfrag{t}{t}
\psfrag{Pi}[][][1][180]{$\langle \Pi(t) \rangle$}
\psfrag{0.0}{0.0}
\psfrag{0.5}{0.5}
\psfrag{-0.5}{-0.5}
\psfrag{1.0}{1.0}
\psfrag{0}{0}
\psfrag{500}{500}
\psfrag{1000}{1000}
\includegraphics[totalheight=4.0cm]{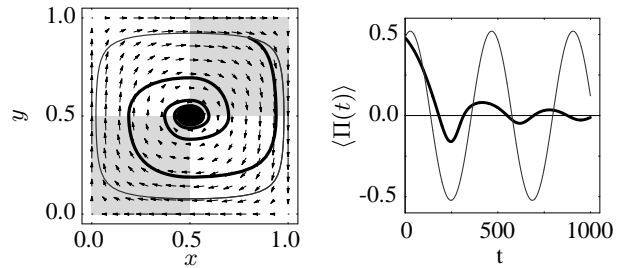}
\caption{Matching pennies:
Comparison between the behavior of a
population with constant learning rate,
i.e.\ $\alpha_x =0$, (thin lines)  and a
population with adaptive learning rate
(perception ability $\alpha_x=10$,
thick lines). The opponent has in both cases
a constant learning rate $\eta_y=1$.
Left: Trajectories in strategy space.
Arrows show the vector field of the
replicator dynamics. Population $X$ has
positive (negative) average payoff in gray
(white) areas. Right: Time development
of the average payoff of the population $X$.
The adaptive learning rate increases the time
intervals in which the corresponding population
has a positive payoff, dampening the
oscillations around the Nash equilibrium
\cite{symp}.}
\label{PM_timedevelopment}
\end{figure}

If we include adaptive learning rates
(\ref{mrule}) into the system, we find
$\dot H(x,y) = -2 \tanh (\alpha_x \Delta \Pi) \Delta \Pi \leq 0$,
vanishing for $\alpha_x=0$. Hence, adaptive
learning rates dampen the oscillations
around the Nash equilibrium and
the trajectories in the $x-y$ plane
spiral towards the Nash equilibrium where
$\langle \Pi_x \rangle=\langle \Pi_y \rangle=0$,
see Fig.\ \ref{PM_timedevelopment}. In addition,
this leads to an increased payoff of one population.
As the matrices (\ref{PMpayoff}) describe
a zero sum game it is sufficient for a population
if it knows its own current average payoff
$\langle \Delta \Pi \rangle =2 \langle  \Pi_x \rangle$.

Numerical simulations for $\alpha_x>0$
show that the temporal integral of the payoff
becomes
\begin{equation}
 \langle \int_{t_0}^{t_1} \langle \Pi_x \rangle dt \rangle_{(x_0,y_0)} =
-\frac{1}{8}\left(H(x_1,y_1)-H(x_0,y_0) \right).
\end{equation}
The averaged initial value $H(x_0,y_0)$ can
be calculated as
$  \int\!\!\!\int_0^1 dx_0  dy_0 H(x_0,y_0) = 8$.
For $t \to \infty$ the system relaxes to the
Nash equilibrium where $H=8 \ln 2$. Hence, we find
for the average cumulated payoff
$\langle \int_{t_0}^{\infty} \langle \Pi_x \rangle dt \rangle_{(x_0,y_0)}  \leq  -\frac{1}{8}\left(8 \ln 2 - 8 \right) \approx 0.307$.
Numerical simulations yield $0.308 \pm 0.005$
independent of $\alpha$. We conclude that a
population can increase its average payoff
if it has an adaptive learning rate
$\alpha_x>0$ and if the game does not
start in the Nash equilibrium. The
adaptation parameter $\alpha$ influences
only the time scale on which the Nash equilibrium
is approached.

Small noise intensities drive the system
away from the fixed point and the population
with the adaptive learning rate gains an increased
payoff. If the noise amplitude $\sigma$ becomes
too large the trajectories will be smeared out
homogeneously over the positive (gray) and
negative (white) payoff regions in phase space
(Fig.\ \ref{PM_timedevelopment}).
This implies that the average gain of population
one decreases to zero. Although the average payoff
is very small even for the optimal noise intensity,
the cumulated payoff increases linearly
in time. This means that for long times the
gained payoff accumulates to a profitable value.

\begin{figure}[htbp]
\psfrag{0}{0}
\psfrag{1}{1}
\psfrag{2}{2}
\psfrag{3}{3}
\psfrag{4}{4}
\psfrag{a1x0}{$\alpha_x=1.0$}
\psfrag{a0x1}{$\alpha_x=0.1$}
\psfrag{sigma}{$\sigma$}
\psfrag{Pi}[][][1][180]{$\langle \Pi \rangle \times 10^{4}$}
\includegraphics[totalheight=4.5cm]{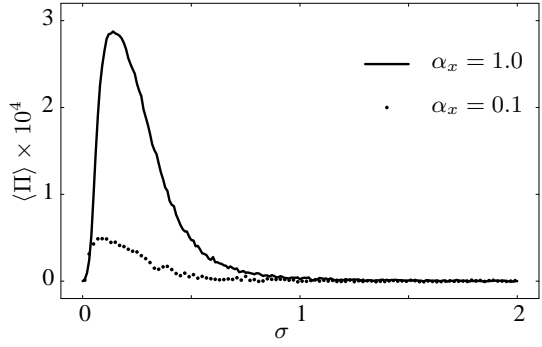}
\caption{Matching pennies:
Average payoff of a population with adaptive
learning rate against a population
with constant learning rate under the
influence of noise for different
noise intensities ($\alpha_y=0$, averages over
$2 \times 10^4$ initial conditions
and $2 \times 10^4$ time steps, see
\cite{symp} for further details).}
\label{PM_noisepayoff}
\normalsize
\end{figure}

As a second application we analyze
the effect of adaptive learning rates
and noise on the prisoner's dilemma.
We use the standard payoff matrix
\cite{Axelrod1984},
\begin{equation}
\label{PDpayoff}
P_x=\left( \begin{array}{cc} 3 & 0 \\ 5 & 1 \\ \end{array} \right)=P_y,
\end{equation}
where rows and columns are placed
in the order ``cooperate'', ``defect''.
As this game is not a zero sum game,
the population with the adaptive
learning rate must be able to
compare its own average payoff
with the opponent's average payoff.
The replicator dynamics of this
system is determined by Eqs.\
(\ref{RD}) and (\ref{PDpayoff}),
\begin{eqnarray}
\label{PDRD}
\dot x &=& x \eta_x (x-1)(1+y) +\xi_x \\ \nonumber
\dot y &=& y \eta_y (y-1)(1+x) + \xi_y.
\end{eqnarray}
There is a stable fixed point in the
Nash equilibrium $x=y=0$ where
both players defect and an unstable
fixed point for mutual cooperation,
 i.e., $x=y=1$.

\begin{figure}[htbp]
\psfrag{0}{0}
\psfrag{1}{1}
\psfrag{2}{2}
\psfrag{0.00}{0.00}
\psfrag{0.05}{0.05}
\psfrag{-0.05}{-0.05}
\psfrag{0.10}{0.10}
\psfrag{0.15}{0.15}
\psfrag{a1x0}{$\alpha_x=1.0$}
\psfrag{a0x1}{$\alpha_x=0.1$}
\psfrag{sigma}{$\sigma$}
\psfrag{Pi}[][][1][180]{$ \langle \Delta \Pi \rangle$}
\includegraphics[totalheight=4.5cm]{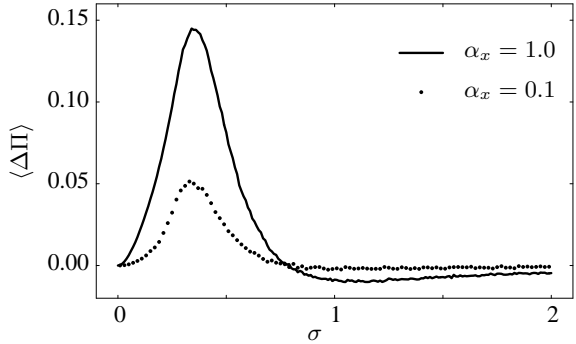}
\caption{Prisoner's dilemma:
Average payoff difference of a population with adaptive
learning rate against a population
with constant learning rate for different
noise intensities.
The negative payoffs arise from the fact that
we have $\eta_x<\eta_y$ for $x<y$
($\Delta t=0.01$, $\alpha_y=0$,
averages over
$2\times 10^4$ initial conditions and
$2 \times 10^4$ time steps).}
\label{PD_noisepayoff}
\end{figure}

The average payoff difference under the influence
of noise is similar as in matching pennies.
Small fluctuations lead the system slowly away
from the Nash equilibrium and tend to increase the payoff.
If the fluctuations are too large they disturb population with
adaptive learning rates and the payoff decreases again,
see Fig.\ \ref{PD_noisepayoff}. Interestingly
enough, here too much noise even leads to
a decreasing payoff difference.

In order to describe the ``stochastic gain'' effect
analytically we introduce
a simplified model. A linearization 
of Eq.\ (\ref{PDRD}) around the 
stable Nash equilibrium leads 
for constant learning rates
to $\dot x = -\eta_x x +\xi_x$ and 
$\dot y = - \eta_y y+\xi_y$. We now
analyze a game in which the replicator 
dynamics is given by these linear equations
and include adaptive learning rates based on
the payoffs for the prisoner's dilemma. 
With $\Delta \Pi = -5(x-y)$ the adaptive 
learning rate $\eta_x$ becomes
$\eta_x=1+\tanh(5\alpha(x-y)) \approx 1 + 5 \alpha(x-y)$
for $\alpha, x,y \ll 1$. The simplified
system can be viewed as a small noise 
expansion of the prisoner's dilemma, 
where the trajectory stays close to the
Nash equilibrium. For $\eta_y=1$ the 
simplified noisy replicator equations
read
\begin{subequations}
\begin{eqnarray}
\label{linPDRDx}
\dot x & = & - x - \alpha' x (x- y) + \xi_x  \\
\label{linPDRDy}
\dot y & =&  - y + \xi_y,
\end{eqnarray}
\end{subequations}
where $\alpha'=5 \alpha$.
The effect of different
constant learning rates is discussed in \cite{Bergstrom2003}.
The mechanism we introduce here is more intricated,
as the adaptive learning rate leads to a dynamical 
adjustment of the learning rate and
the average of $\eta_x = 1+ \alpha'(x-y)$ over
all possible strategies is $\eta_y=1$. 

\begin{figure}[htbp]
\psfrag{sigma}{$\sigma$}
\psfrag{gain}[][][1][180]{$\langle \Delta \Pi \rangle$}
\psfrag{0.00}{0.00}
\psfrag{0.01}{0.01}
\psfrag{sim}{simulation}
\psfrag{ana}{analytical curve}
\psfrag{0}{0}
\psfrag{1}{1}
\psfrag{2}{2}
\includegraphics[totalheight=4.5cm]{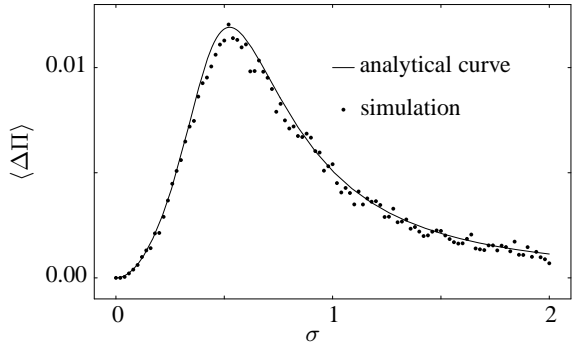}
\caption{Simplified model: Comparison
of the average payoff difference
$\langle \Delta \Pi \rangle$ 
from a simulation of Eqs.\
(\ref{linPDRDx},\ref{linPDRDy})
and the analytical function Eq.\
(\ref{analytPay})  ($\Delta t = 0.01$,
$\alpha'=5 \alpha=0.1$, averages over
$4 \times 10^4$ time steps and
$4 \times 10^4$ realizations).}
\label{SPD}
\end{figure}

Equation (\ref{linPDRDy}) describes an
Ornstein-Uhlenbeck process \cite{Uhlenbeck1930}, 
here the dynamics is restricted to $0 \leq y \leq 1$.
The Fokker-Planck equation \cite{Gardiner1985}
for $p_y=p_y(y,t|y_0,t_0)$,
\begin{equation}
\dot p_y = \frac{d}{dy} \left(  y p_y + \frac{\sigma^2}{2} \frac{d}{dy} p_y \right),
\end{equation}
has the stationary solution
$p_y^s= {\cal N}_y e^{-y^2/\sigma^2}$,
where ${\cal N}_y^{-1}=\int_0^1 e^{-y^2/\sigma^2} dy$.
We find the mean value
$\langle y(\sigma) \rangle$ as
\begin{equation}
\label{ym}
\langle y \rangle = \int_0^1 dy p_y y = 
\frac{ \sigma \left( 1 -    e^{-{\sigma }^{-2}}     \right) }{{\sqrt{\pi }}   {\rm Erf}(\frac{1}{\sigma })}. 
\end{equation}
$y$ is a correlated stochastic process 
which appears in Eq.\ 
(\ref{linPDRDx}) as a multiplicative noise. 
Numerical simulations indicate that we may 
neglect the stochastic nature of y and replace it 
by $\langle y \rangle$ for small $\alpha$.
This leads to an approximated Fokker-Planck 
equation for  $p_x=p_x(x,t|x_0,0)$ 
\begin{equation}
\dot p_x = \frac{d}{dx} \left[ -a(x) p_x + \frac{\sigma^2}{2} \frac{d}{dx} p_x  \right] 
\end{equation}
where
$a(x)  =  - x - x \alpha' (x - \langle y \rangle)$.
Since $x$ is (similarly to $y$)
also restricted to $ 0 \leq x \leq 1$
we find the stationary solution
\begin{equation}
\label{pxeq}
p_x^s = {\cal N}_x \exp \left[  -\frac{x^2}{\sigma^2} -\frac{2 \alpha' x^3}{3 \sigma^2}+\frac{\alpha' \langle y \rangle x^2}{\sigma^2}\right]
\end{equation}
with the normalization constant ${\cal N}_x$. 
Since $x$ is typically of the order of $\sigma$
for $\sigma \ll 1$ the term $x^2/\sigma^2$ is finite.
Therefore we can expand Eq.\ (\ref{pxeq}) for $\alpha' \ll 1$
and obtain expanding $\langle x \rangle$ again
an analytical expression for 
$\langle \Delta \Pi \rangle = -5 (\langle x \rangle-\langle y \rangle)$ 
\begin{eqnarray}
\label{analytPay}
\langle \Delta \Pi \rangle & = & - 5 \alpha' \frac{d}{d \alpha'} \langle x \rangle
=5 \alpha' \Bigg[ \frac{\sigma^2}{2} 
- \delta^3  \sigma \gamma (1-\gamma)^2 \\ \nonumber
& + & \delta^2 (1-\gamma) \left( \frac{5}{3} \gamma  - \frac{7}{6} \sigma^2 (1-\gamma)\right)
 - \delta \gamma \left(\frac{2}{3 \sigma}+\sigma \right)  \Bigg],
\end{eqnarray}
where 
$\delta = \frac{1}{\sqrt{\pi}{\rm Erf}(1/\sigma)}$ and $\gamma = e^{-1/\sigma^{2}}$. 
The asymptotics of Eq.\ (\ref{analytPay}) can be computed as 
$\langle \Delta \Pi \rangle = \alpha'/\left(24 \sigma^2\right)$ for $\sigma \gg 1 $ 
and $\langle \Delta \Pi \rangle = \alpha' \left(\frac{5}{2} -\frac{35}{6 \pi} \right)\sigma^2$
for $\sigma \ll 1$.
We stress that this simplified system which consists of 
a stable fixed point with linear adaptive learning rate 
in the presence of noise is the simplest possible model
that describes the stochastic gain effect. 
Fig.\ \ref{SPD} shows a comparison between 
the analytical payoff difference Eq. (\ref{analytPay})
and a simulation of Eqs.\ (\ref{linPDRDx},\ref{linPDRDy}).

To summarize, we have introduced an extension
to the usual replicator dynamics
that modifies the the learning rates
using a simple ``win stay---lose shift''
rule. In this way, a population optimizes
the payoff difference to a competing population.
This simple rule leads to a convergence
towards the mixed Nash equilibrium
for the game of ``matching pennies'' \cite{RPS}.
Even in games with stable Nash equilibria
as the ``prisoner's dilemma''
transient phases can be exploited,
although the basins of attraction are not altered,
as e.g.\ in \cite{Bergstrom2003}.
Weak external noise drives the system into
the transient regime and leads to an increased
gain for one adaptive population.

In conclusion, we have found a learning process
which improves the gain of the population with
adaptive learning rate under the influence of external noise.
Fluctuations lead to an increased payoff for
intermediate noise intensities
in a resonance-like fashion.
This phenomenon could be of particular
interest in economics, where
interactions are always subject to external 
disturbances \cite{Cabrales2000,Fudenberg1990,Gintis2000}.

\acknowledgments{
We thank J. C. Claussen for
stimulating discussions and
comments on this manuscript.
A.T.\ acknowledges support
by the Studienstiftung des deutschen Volkes.}

\bibliographystyle{apsrev}

\begin{thebibliography}{99}
\bibitem{RPS}			A comparison with ``rock-paper-scissors'' reveals
				that this is not necessarily the case, as there are periodic
				attractors with $\Delta \Pi=0$ leading to constant $\eta_x$. However,
				for the chaotic extension of this game \cite{Sato2002a} these attractors
				vanish and the trajectory converges to the Nash equilibrium, as expected by Sato {\em et al.}
				\cite{Sato2002a}.
\end{thebibliography}

\end{document}